\documentclass[twocolumn,astrosymb, times]{aastex701}
\usepackage{rotating}
\usepackage{booktabs}
\usepackage{textcomp, gensymb}
\usepackage{tabularx}
\usepackage{amsmath}

\shortauthors{P\'erez Paolino et al.}

\begin{document}

\title{Separating Photospheric and Starspot Magnetic Fields in Pre-Main-Sequence Stars Using IGRINS Spectroscopy}
\correspondingauthor{Facundo P\'erez Paolino}
\author[0000-0002-4128-7867, gname=Facundo, sname=P\'erez Paolino]{Facundo P\'erez Paolino}
\affiliation{Department of Astronomy, California Institute of Technology, 1216 East California Blvd, Pasadena, CA 91125, USA}
\email[show]{fperezpa@caltech.edu}

\author{Lynne A. Hillenbrand}
\affiliation{Department of Astronomy, California Institute of Technology, 1216 East California Blvd, Pasadena, CA 91125, USA}
\email{lah@astro.caltech.edu}

\author[0000-0001-8642-5867]{Jeffrey S. Bary}
\email{jbary@colgate.edu}
\affiliation{Colgate University, 13 Oak Drive, Hamilton, NY 13346, USA}



\begin{abstract}
Magnetic fields in pre-main-sequence stars regulate angular momentum evolution, drive magnetic activity, and modify stellar structure, yet their surface distributions remain poorly constrained. Traditional single-component Zeeman broadening analyses typically yield mean field strengths of 2–4 kG, sometimes exceeding the photospheric equipartition limit, and assume complete magnetic coverage. These assumptions conflict with evidence that strong fields are concentrated in cool starspots. Here we present the first systematic separation of photospheric and starspot magnetic field strengths in PMS stars, using high-resolution ($R \approx 45{,}000$) H- and K-band spectra from the Raw and Reduced IGRINS Archive. By modeling temperature and magnetic field strength simultaneously for a vetted sample of 33 Class II–III young stellar objects, we find median photospheric field strengths of 1.2 kG and median spot field strengths over two times stronger at 2.56 kG, resolving the apparent super-equipartition tension and removing the need for a unity magnetic filling factor. Our results show that PMS surfaces are permeated by concentrated, kG-strength spot fields covering 27–83\% of the visible hemisphere. This two-component framework offers a physically motivated means to reconcile spectroscopic and imaging-based magnetic diagnostics and enables large-scale magnetic population studies across young clusters and star-forming regions.
\end{abstract}


\section{Introduction}
Pre-main-sequence (PMS) stars possess strong magnetic fields (on the order of several kiloGauss, kG) that influence nearly every aspect of their early evolution \citep{Basri2021book}. From regulating angular momentum to mediating mass accretion to producing chromospheric activity and surface inhomogeneities, magnetic fields play a central role in shaping young stars during their first tens of millions of years \citep{Donati2009}. Recent studies have also demonstrated that magnetic fields impact stellar interiors, leading to radius inflation, suppressed convection, altered effective temperatures, and deviations from standard pre-main-sequence evolutionary tracks \citep{Somers2015, Feiden2016,Cao2025}.

Lines with large Land\'e $g$-factors exhibit excess broadening beyond what is expected from thermal and rotational effects, owing to Zeeman splitting of the energy levels. This phenomenon, known as \textit{Zeeman Broadening (ZB)}, scales as $\lambda^{2}$, making near-infrared spectra ideal for measuring magnetic fields from the ground. In practice, however, ZB requires relatively strong surface magnetic fields and modest projected rotational velocities for the splitting to remain detectable above other line-broadening mechanisms \citep[e.g.,][]{yang_magnetic_2008}. A complementary effect is \textit{Zeeman Intensification (ZI)}, in which magnetic splitting redistributes the oscillator strength of the transition into multiple components, desaturating the line core and enhancing the line’s equivalent width. This increase in absorption at wavelengths offset from the line center effectively raises the line opacity across a broader frequency range \citep{Basri1992}. {Unlike ZB, ZI is less sensitive to high signal-to-noise requirements and to rotational broadening, since it relies on the integrated equivalent width rather than the resolved profile \citep{han_magnetic_2023}.}

Persistent puzzles in the field concern the interpretation of these strong surface-averaged magnetic fields. Single-component fits to Zeeman-broadened lines in PMS stars often yield mean field strengths of 2–4 kG \citep[e.g.,][]{johnskrull_magnetic_2007, Shulyak2019}, with some reported values exceeding the maximum magnetic pressure balanced by gas pressure in the photosphere or the equipartition field strength \citep[e.g.,][]{Safier1999}. These “super-equipartition” fields have raised concerns about the physical plausibility of such large field strengths and the limitations of single-temperature, single-field models. Furthermore, many such models implicitly assume a magnetic filling factor of unity, which likely overestimates the spatial extent of the field and under-represents the inhomogeneous, spot-dominated surfaces of most PMS stars. 

Starspots arise where concentrated strong magnetic fields inhibit convective energy transport, producing large, cool regions on the stellar surface \citep{Strassmeier2009}. As such, they serve as observable tracers of magnetic activity. The presence of starspots in young stars is now well established, supported by multiple lines of evidence: periodic light curve modulation due to rotationally-modulated spot visibility \citep{Vrba1986, herbst1990}, magnetic field mapping via Zeeman-Doppler-Imaging (ZDI) (e.g., \citealt{Donati2019}), spectroscopic indicators such as spectral type mismatches with wavelength \citep[e.g.][]{Vacca2011}, and multi-component spectral modeling with distinct photospheric and spot temperatures \citep{Debes2013, Bary2014, Gully2017, Cao2022, PerezPaolino2024}. ZDI results have indicated the presence of (large$\geq20\%$ of the stellar surface) starspots with strong surface magnetic fields, but fail to capture the integrated magnetic field strength from small scale features, representing only at best a fourth of the total magnetic flux \citep{See2020}. 

Previous spectroscopic studies of spotted stars combined two atmospheres of different temperatures, the starspots and the photosphere, covering some fraction of the stellar surface to construct their spectra \citep[e.g.,][]{Gully2017}. {These models typically model the photoshophere as being isothermal, with starspots covering some fraction of the stellar surface emitting at a lower, also isothermal temperature. While this is at best a crude oversimplification of the likely temperature gradients between starspots and photosphere, these models still provide useful estimates of the bulk surface averaged spot coverage and temperature.} Here, we modify this approach to allow each component to have its own magnetic field strength. Such an approach combines the sensitivity of the spectral features in the $K$-band to the strength of the B-field with the temperature sensitivity of features in the $H$-band, allowing $T_{spot}$, $T_{phot}$, $B_{spot}$ and $B_{phot}$ be constrained from the same observation when using a high-resolution cross-dispersed spectrometer. We use a set of magnetic theoretical atmospheres computed using the MOOGSTOKES magnetized one-dimensional LTE radiative transfer code \citep{Deen2013} to leverage the effects of ZI and ZB in our fits. In this work, we apply this methodology to a carefully selected sample of 34 Class II and III PMS stars observed with IGRINS at R~$\approx$~45000, providing the first systematic measurements of separate photospheric and starspot field strengths. Our analysis resolves long-standing tensions between spectroscopic and imaging-based magnetic diagnostics and offers a physically consistent framework for interpreting magnetic activity in PMS stars. {While 1D LTE radiative transfer is an oversimplification of the inherently three-dimensional, non-LTE nature of starspot and photospheric structure, they allow for direct comparison with prior work, and, crucially, reproduce the bulk line-depth and Zeeman broadening signatures at the spectral resolution of IGRINS. In this sense, they provide an appropriate framework for deriving surface-averaged field strengths and spot parameters, even if they cannot capture fine-scale gradients or dynamics.}

\section{Data and Sample Selection}\label{sec:sample_and_obs}
This study makes use of publicly available high-resolution near-infrared spectra from the Immersion GRating INfrared Spectrometer\footnote{\url{https://github.com/igrins/plp}} (IGRINS; \citealt{Park2014}) obtained through the Raw and Reduced IGRINS Spectral Archive (RRISA; \citealt{RRISA}), which hosts uniformly reduced and telluric-corrected spectra. The IGRINS instrument provides simultaneous coverage of the entire $H$ and $K$-bands (1.45-2.5 µm) at a resolving power of R$\approx$45000, with high wavelength stability and well-characterized instrumental broadening.

We used the RRISA archive to identify spectra of Class II and III PMS stars. For this, we used the samples of \citet{lopez2021}, \citet{lopez-valdivia_igrins_2023}, \citet{PerezPaolino2024}, and \citet{PerezPaolino2025b}, resulting in a sample of $\sim$~200 stars. We then made a cut for stars possessing a signal-to-noise ratio (S/N) greater than 100 in the $K$-band and that were free of strong accretion signatures and veiling between 2.20 and 2.315~\micron. {To this end, we enforced a cutoff in K-band veiling to $r_K<3$ and looked for stars lacking obvious Brackett or CO emission.} The resulting sample consisted of $\sim$104 stars with spectra in RRISA. Of those sources, we prioritized ones for which multi-component spectroscopic spot models exist in the literature (e.g., \citealt{Gangi, PerezPaolino2024}) if possible. In cases where multiple observations of a star were available, we used the data with highest S/N. All spectra were visually inspected to ensure data quality, proper telluric correction, and absence of significant contamination from instrumental artifacts. From the $\sim104$ PMS stars we analyzed, our final sample consists of the 33 stars Class II and III PMS shown in Table~\ref{tab:obs} for which our fits with starspots and separate magnetic components showed strong statistical significance (See \ref{sec:models}). We expect the remaining 71 stars to be spotted, but we could not separate the components with our H- and K-band spectra. All of these were successfully fit with single-temperature, single-magnetic component models however, and will be presented in future work.  

\section{Models}\label{sec:models}
\subsection{Magnetic Models of Spotted Stars}
We used the grid of theoretical atmospheres computed using MOOGSTOKES of \citet{LopezValdiviaCode}. In MOOGSTOKES, the emergent spectrum of a star is computed incorporating the Zeeman splitting from a purely radial magnetic field. As a basis for the MOOG calculations, we used MARCS synthetic atmospheres \citep{Gustafsson2008} with solar metallicity. The line-list used in the calculation is from the Vienna Atomic Line Database (VALD; \citealt{Ryabchikova2015}), modified with the updated Van der Waals constants and oscillator strengths of \citet{Flores2019}. This method follows the approach of \citet{lopez2021} and we direct the reader there for a full discussion of the radiative transfer code.  The final grid of theoretical atmospheres spans the parameter space $T_{\mathrm{eff}} \in [3000, 5000]~\mathrm{K}$, sampled in steps of $100~\mathrm{K}$ between $3000$ and $4000~\mathrm{K}$, and in steps of $250~\mathrm{K}$ thereafter; $\log g \in [3.0, 5.0]$~dex in steps of $0.5$~dex; $v\sin i \in [2, 50]~\mathrm{km~s^{-1}}$ in steps of $2~\mathrm{km~s^{-1}}$; and $B \in [0, 4]~\mathrm{kG}$ in steps of $0.5~\mathrm{kG}$. The grid of synthetic spectra was then linearly interpolated  across all parameters to generate a continuous sequence of theoretical stellar atmospheres.

\subsection{Fitting Procedure}
We follow the methodology of \citet{PerezPaolino2025b} to construct two-component models of spotted stars in the form 

\begin{equation}
    F_{\lambda,\ \mathrm{model}} = F_{\lambda}(T_{\mathrm{phot}})\,(1-f_{\mathrm{spot}}) + F_{\lambda}(T_{\mathrm{spot}})\,f_{\mathrm{spot}}
\end{equation}

\noindent
where $F_{\lambda}(T_{\mathrm{phot}})\,(1-f_{\mathrm{spot}})$ is the photospheric flux density contribution and $F_{\lambda}(T_{\mathrm{spot}})\,f_{\mathrm{spot}}$ is the starspot flux density contribution, weighted by the spot filling factor $f_{\mathrm{spot}}$. However, we allow for two separate magnetic components, $\rm B_{phot}$ and $\rm B_{spot}$, during the fitting process. Excess continuum emission weakens spectral lines in the H and K-bands as the additional continuum emission fills in or veils the absorption lines \citep[e.g.,][]{Hartigan1991}. Therefore, we veil our synthetic spectra following 

\begin{equation}
    F_\lambda = \frac{F_{\lambda}+r_{H/K}}{1+r_{H/K}}
\end{equation}

\noindent
where $r_{H/K}$ is the veiling at H- or K-band. {Here, veiling is treated as being constant across wavelengths. This is a rough approximation, but a more thorough treatment of wavelength-dependent veiling is beyond the scope of this work \citep[e.g.,][]{Dodin2012}.} Each model is then determined by a combination of ten parameters: $\rm f_{spot}$, $\rm T_{phot}$, $\rm T_{spot}$, log~$g$, $v\ sini$, $\rm v_{radial}$, $\rm B_{phot}$, $\rm B_{spot}$, $\rm r_H$, and $\rm r_K$.  These ten parameters are fit simultaneously during the fitting process outlined below.

The $K$-band has been extensively employed to measure magnetic fields in young stars due to its high magnetic sensitivity \citep[see, e.g.,][]{johnskrull_magnetic_2007}. \citet{Flores2022} measured magnetic fields for a sample of 40 Class II and III YSOs using iShell high resolution (R$\sim$50000) spectra, finding magnetic fields on the order of 2-3~kG. In a followup study, \citet{flores_ishell_2024} used the same methodology to study a sample of Class I and Flat Spectrum (FS) sources, finding no statistically significant difference in the magnetic field strengths across both samples. These results highlight the usefulness of K-band high resolution spectroscopy for determining magnetic field strengths and stellar parameters. However, it lacks atomic and molecular features that strongly constrain spot temperatures \citep[see, e.g.,][]{PerezPaolino2024}. To address this limitation, we supplemented the 2.200–2.315~\micron\ region of the $K$-band with the 1.558–1.568~\micron\ region of the $H$-band. We are limited to these regions by the spectral coverage of the theoretical atmosphere grid. However, this $H$-band window contains temperature-sensitive lines, including atomic Fe\,\textsc{i} at 1.56259 and 1.56362~\micron, as well as molecular OH at 1.56310 and 1.56317~\micron. These features were used by \citet{Tang2024} to develop an empirical temperature relation, owing to their inverse temperature sensitivity over the 3000-5000~K temperature range studied here. 

\begin{figure*}
\centering
\includegraphics[width=0.9\linewidth]{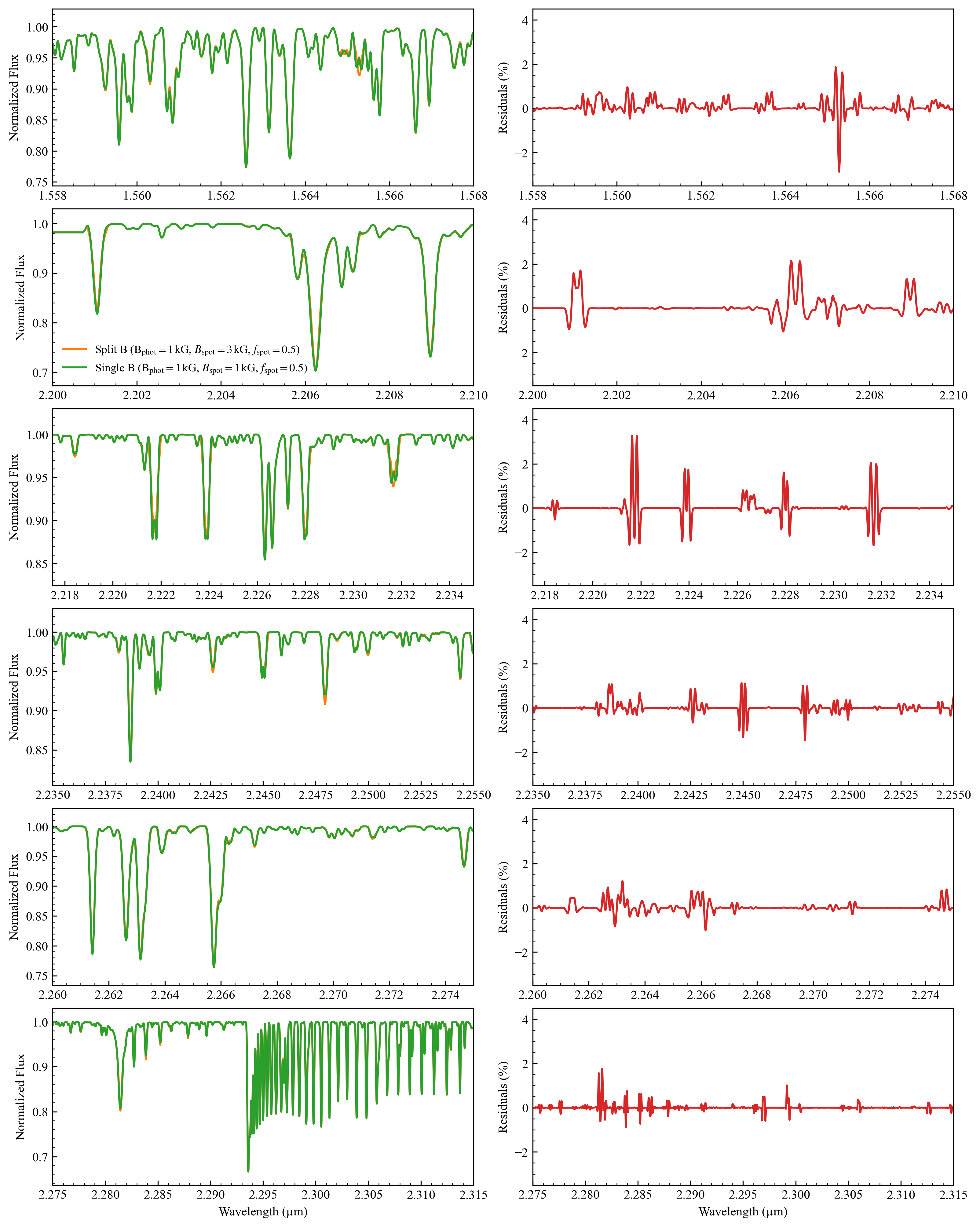}
\caption{{Comparison of synthetic spectra for a two-field configuration to a single-field configuration for the six wavelength regions used in fitting, each with the same mean magnetic flux and projected rotational velocity $v \sin i = 10~\mathrm{km\,s^{-1}}$.} The two-field model has $B_{\mathrm{phot}} = 1~\mathrm{kG}$ and $B_{\mathrm{spot}} = 3~\mathrm{kG}$ with $f_{\mathrm{spot}} = 0.5$, while the single-field model has $B_{\mathrm{phot}} = B_{\mathrm{spot}} = 2~\mathrm{kG}$ and $f_{\mathrm{spot}} = 0.5$. Both models have $T_{\rm phot}=4250~K$ and $T_{\rm spot}=3600~K$. Left panels show the normalized flux in each wavelength interval; right panels show the fractional residuals between the two models. Residual amplitudes of $1$--$4\%$ occur across multiple magnetically sensitive lines, indicating that high-S/N ($\gtrsim 100$) spectra should permit detection of such differences in real data.}\label{fig:comp}
\end{figure*}

\begin{figure*}
\centering
\includegraphics[width=0.9\linewidth]{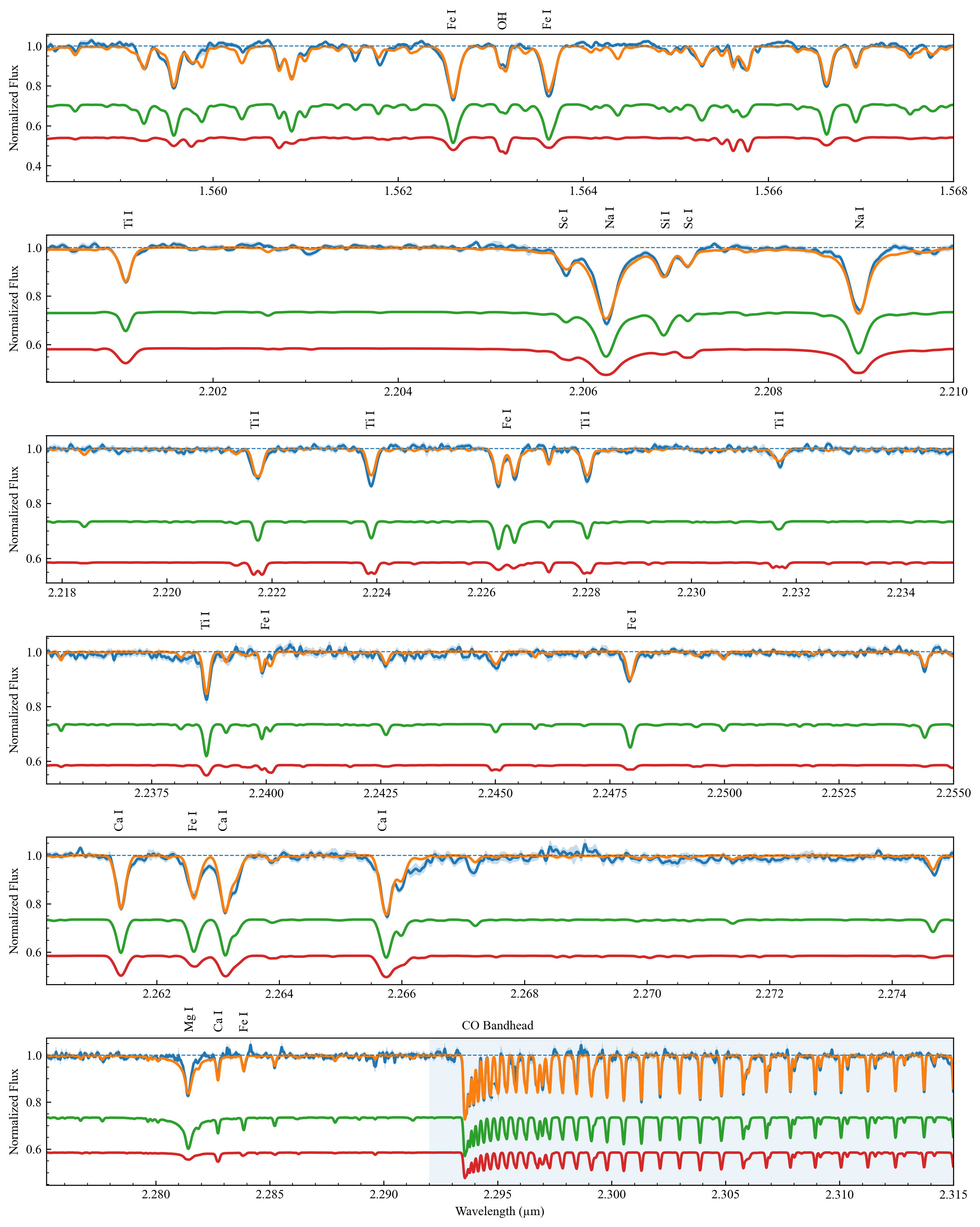}
\caption{{Comparison of IGRINS spectra (blue) to best-fitting two-component spot models (orange) for Haro~1-8 across selected wavelength regions.} The total model (orange) is decomposed into the starspots (red) and the unspotted photosphere (green). A horizontal dashed line marks the normalized continuum level. Strong lines are marked throughout, with the CO Bandhead region shaded in light blue.}\label{fig:fit}
\end{figure*}

The final set of regions used for fitting is shown in Figure~\ref{fig:comp} and Figure~\ref{fig:fit}. The top panel shows the region containing the Fe\,\textsc{i} and OH features, while the rest overlap those used by \citet{lopez2021} and \citet{Flores2019}. These regions include the magnetically sensitive Na\,\textsc{i} doublet at 2.2062 and 2.2090~\micron, Sc\,\textsc{i} lines at 2.2058 and 2.2071~\micron, and a Si\,\textsc{i} line at 2.2068~\micron\ that are more sensitive to $T_{\rm eff}$. Also present are Ti\,\textsc{i} lines at 2.2217, 2.2239, and 2.2315~\micron\ that are highly magnetically sensitive, as well as Fe\,\textsc{i} lines primarily sensitive to $T_{\rm eff}$. The Ca\,\textsc{i} lines at 2.2614, 2.2631, and 2.2657~\micron, together with an Fe\,\textsc{i} line at 2.2626~\micron, are most useful for constraining $T_{\rm eff}$ and $\log g$ because their magnetic sensitivity is weak. Finally, the region from 2.2986–2.3150~\micron\ contains multiple CO Band lines that are sensitive to $\log g$, which strengthen as $\log g$ decreases.

Spectra were continuum-subtracted using an asymmetric least squares algorithm (see e.g., \citealt{Eilers}) prior to fitting. Fits were performed using the \texttt{emcee} Markov Chain Monte Carlo sampler \citep{Foreman-Mackey2013}, adopting a chi-squared likelihood function

\begin{equation}
    \ln p = -\frac{1}{2} \sum_n \left[ 
        \frac{(y_{\mathrm{data}} - y_{\mathrm{model}})^2}{\sigma^2} + \ln \left( 2\pi \sigma^2 \right) 
    \right]
\end{equation}

\noindent
where $\sigma$ is the per-pixel uncertainty and $y_{\mathrm{data}}$ and $y_{\mathrm{model}}$ are the observed and model fluxes, respectively. {We do not adopt the raw instrumental uncertainty, $\sigma_{\mathrm{data}}$, directly. Instead, following the approach of \citet{Robinson2019} we add in quadrature a constant fractional uncertainty to account for model imperfections and residual instrumental systematics not captured by the pipeline error arrays. We experimented with values between 0.5–2\% and found that 1\% was the smallest addition that consistently brought the distribution of reduced chi-squared values for well-constrained fits close to unity}. Therefore we use an effective uncertainty per pixel of  
\begin{equation}
    \sigma_{\mathrm{eff}} = \sqrt{\sigma_{\mathrm{data}}^2 + \left( y_{\mathrm{model}}\, k_{\mathrm{unc}} \right)^2}
\end{equation}
where $k_{\mathrm{unc}} = 0.01$. We used 50 walkers and ran them for 5000 steps each, discarding the first 2500 steps. This was sufficient to achieve convergence in all fits, as was confirmed through visual inspection of each individual chain. An example posterior distribution is shown in Figure~\ref{fig:posterior}, with overlaid medians and $3\sigma$ percentiles. Most of the fit parameters returned symmetric posteriors, which we report as $3\sigma$ uncertainties. In cases where the posteriors were asymmetric, we used the larger $3\sigma$ uncertainty.

Finally, we calculate spot-corrected effective temperatures for all the stars in our sample using
\begin{equation}\label{eq:teff}
    T_{\rm eff}=[T_{\rm phot}^4(1-f_{\rm spot})+T_{\rm spot}^4f_{\rm spot}]^{0.25}
\end{equation}
and report these values in the last column of Table~\ref{tab:obs}.

\subsection{Sensitivity of IGRINS Spectra to Two Magnetic Components}
Detection limits for magnetic fields using IGRINS spectra have been well established. \citet{han_magnetic_2023} report detection thresholds of $\approx0.7$~kG for M-dwarfs with S/N~$\geq$~150 spectra using the $ZI$ method, which is more sensitive than for the $ZB$ method at 0.87~kG. An additional advantage is that, although rapid rotation-rates can introduce degeneracy with magnetic broadening, this effect is more severe for the ZB method. For ZI, \citet{Hussaini2020} estimate that rotational broadening sets upper bounds on detectability of B(kG)$\sim v\sin i/10$~km~s$^{-1}$ for Ti~lines and B(kG)$\sim v\sin i/5$~km~s$^{-1}$ for Na lines. By combining tens of lines across multiple spectral regions, our analysis pushes detection limits to B(kG)$v\sin i \gtrsim 10$~km~s$^{-1}$. While most of the stars in our sample have low $v\sin i$ values, there are a few with $v\sin i \gtrsim 20$~km~s$^{-1}$. The B-field measurements for those are inherently more uncertain and should be treated with caution.

To showcase the differences between a spotted--star model with a single magnetic field and one with two separate magnetic components, Figure~\ref{fig:comp} presents an example of both cases. The single--field model ($\rm B_{phot} \approx B_{spot}$) has a strength of 2~kG, while the two--component model has identical temperatures, spot coverage, and the same median field strength of 2~kG, but with $\rm B_{phot}=1$~kG and $\rm B_{spot}=3$~kG. Every other parameter ($v \sin i = 10~\mathrm{km\,s^{-1}}$, $f_{\mathrm{spot}} = 0.5$,  $T_{\mathrm{spot}} = 3600~K$,  $T_{\mathrm{phot}} = 4250~K$, log~$g$=4.0, $\rm r_H=0.5$, and $r_K$=1) is identical between the two models. Differences in the normalized fluxes are on the order of 1-4\% with RMS residuals across all windows of order $0.5-1.0\%$, demonstrating the capability of modeling of high SNR ($S/N\geq100$), high-resolution spectra to separate both magnetic components.

\section{Results}\label{sec:results}
Across the 33 stars in our sample, we measure substantial starspot coverage, with filling factors ranging from 0.27 to 0.83 and a median value of $f_{\mathrm{spot}}=0.55$ ($\sigma=0.14$). These values are in agreement with previous results by \citet{Cao2022, Gangi, perez_paolino_starspots_2025} for stars of similar ages. Photospheric temperatures have a median of $T_{\mathrm{phot}}=4250$ K, with spot temperatures cooler by $\sim$670 K on average ($T_{\mathrm{spot}}=3582$ K), corresponding to a median spot-to-photosphere temperature ratio of 0.84, similarly in good agreement with predictions by \citet{Herbst2020} and empirical measurements from \citet{perez_paolino_starspots_2025}. These values indicate that strong magnetic fields are concentrated within cooler surface regions, which occupy more than half the visible stellar surface in a typical object.

Concerning multiple components of the magnetic field strengths, across the sample we find median fields of $B_{\mathrm{phot}} = 1.20 \pm 0.25~\mathrm{kG}$ and $B_{\mathrm{spot}} = 2.56 \pm 0.61~\mathrm{kG}$. This represents a median $\rm \Delta B=1.36~kG$ measured at a $\sim 8.7\sigma$ difference.

In Figure~\ref{fig:fit} we present an example fit for Haro~1-8. The best-fit model yields a spot filling factor of $f_{\mathrm{spot}} = 0.47 \pm 0.10$, a photospheric temperature of $T_{\mathrm{phot}} = 4467 \pm 94$~K, and a spot temperature of $T_{\mathrm{spot}} = 3716 \pm 89$~K. The surface gravity is $\log g = 4.47 \pm 0.06$~dex, with $v \sin i = 5.19 \pm 0.47$ km s$^{-1}$ and a radial velocity of $v_r = -22.37 \pm 0.14$ km s$^{-1}$. The veiling is $r_H = 0.35 \pm 0.07$ in the $H$-band and $r_K = 0.48 \pm 0.05$ in the $K$-band. The magnetic field strengths are $B_{\mathrm{phot}} = 0.88 \pm 0.19$~kG in the photosphere and $B_{\mathrm{spot}} = 2.08 \pm 0.32$~kG in the spots. This constitutes a large $\rm \Delta B=1.2~kG$ difference in magnetic field strength between the photosphere and the starspots.

{Comparing the Bayesian Information Criterion (BIC) of this model to that of the best-fit single-temperature, single-magnetic-field model yields $\Delta\mathrm{BIC} = 139.33$. Because the BIC includes an explicit penalty for additional free parameters, this large value cannot be attributed to the two-temperature, two-field model simply having greater flexibility. Rather, it provides decisive statistical evidence overwhelmingly favoring the spotted, two-field model despite its added complexity. On the \citet{BICKaas} scale, $\Delta\mathrm{BIC} > 10$ is already considered very strong evidence; a value of 139 corresponds to odds vastly exceeding any reasonable threshold for preference. According to the $\Delta\mathrm{BIC} > 10$ criterion, all of the stars in our sample with the exception of FVTau show a strong preference for the spotted, two magnetic field model. In the case of FVTau, it is the fastest rotator in our sample ($v \sin i = 33.96 \pm 0.86$ km~s$^{-1}$), so the reduced statistical leverage of the model is unsurprising.
}
Since dynamo efficiency and magnetic topology are often expected to vary with stellar structure and temperature \citep[e.g.,][]{Reiners2012, See2020} one would expect a correlation between spot coverage and effective temperature or magnetic field strength. However, there seems to be no correlation: the Pearson correlation coefficients are $r=-0.11$ ($p=0.55$) for $B_{\mathrm{phot}}$ vs.\ $f_{\mathrm{spot}}$ and $r=0.16$ ($p=0.37$) for $B_{\mathrm{spot}}$ vs.\ $f_{\mathrm{spot}}$, with similar results from the Spearman rank test with $\rho=-0.11$ ($p=0.55$) and $\rho=0.09$ ($p=0.62$), respectively. For the coldest stars, the differences between the magnetic field strengths of the photosphere and the spots appear to decrease. However, this trend is suggested by three objects and is by no means conclusive.

\section{Discussion}
\subsection{Comparison with Results from Lower Dispersion Spectra}
Several stars in our sample have been previously modeled for starspots using SpeX \citep{rayner2003}, a medium-resolution ($R \approx 2000$) spectrograph providing simultaneous coverage of the $YJ$-, $H$-, and $K$-bands. These analyses are presented in \citet{PerezPaolino2024, perez_paolino_starspots_2025} for AA~Tau, BP~Tau, DK~Tau, LkCa~4, and IW~Tau. In all cases, significant spot coverage was inferred, with temperature contrasts of 500–800~K and spot filling factors comparable to those derived here (e.g., $\rm f_{spot} = 0.68$ in \citet{perez_paolino_starspots_2025} versus $\rm f_{spot} = 0.73$ in our present fit for AA~Tau, $\rm f_{spot} = 0.84$ in \citet{PerezPaolino2024} versus $\rm f_{spot} = 0.83$ here for LkCa~4).

We find a systematic offset in the absolute temperatures with the $H$- and $K$-band magnetic fits yielding photospheric and spot temperatures that are higher by $\approx 600$~K compared to the SpeX-based values. This offset likely arises because the different spectral ranges probe different atmospheric depths. The distinction in spectral resolution between IGRINS ($R \approx 45{,}000$) and SpeX also plays a role: the SpeX fits are largely constrained by the continuum shape, which are predominantly affected by extinction, and molecular absorption features, whereas the IGRINS fits are dominated by resolved, Zeeman-sensitive line profiles. The latter are formed in deeper layers of the stellar atmosphere and at higher temperatures, whereas molecules only exist in the coldest regions of the stellar atmosphere. In addition, the earlier SpeX analyses assumed a zero magnetic field and fixed value for $\log g$ of 3.5, whereas our present fits explicitly separate $B_{\mathrm{phot}}$, $B_{\mathrm{spot}}$, and $\log g$, which can shift the best-fit temperatures while preserving $f_{\mathrm{spot}}$. Despite these differences, we find excellent agreement ($\Delta T_{eff} < 100$~K) between the spot-corrected temperatures (Equation~\ref{eq:teff}) found here and those determined using moderate resolution SpeX spectra.

The comparison is also subject to limitations arising from the variability in the spot coverage of the observable stellar hemisphere due to rotation \citep{perez_paolino_correlating_2023}, and to the possibility of spot evolution \citep[e.g.,][]{Basri2022} over the longer timescales separating our IGRINS and SpeX observations. T~Tauri spots can evolve on month-to-year timescales. {That the spot filling factors remain consistent within these temporal, spectral, and modeling differences suggests that $f_{\mathrm{spot}}$ is behaving as expected for a physical surface coverage parameter, and provides additional confidence in our fitting procedure. This agreement lends support to using $f_{\mathrm{spot}}$ as an empirical constraint in stellar evolutionary models, even when $T_{\mathrm{phot}}$ and $T_{\mathrm{spot}}$ vary with wavelength or modeling approach.}

\begin{figure}
\centering
\includegraphics[width=1\linewidth]{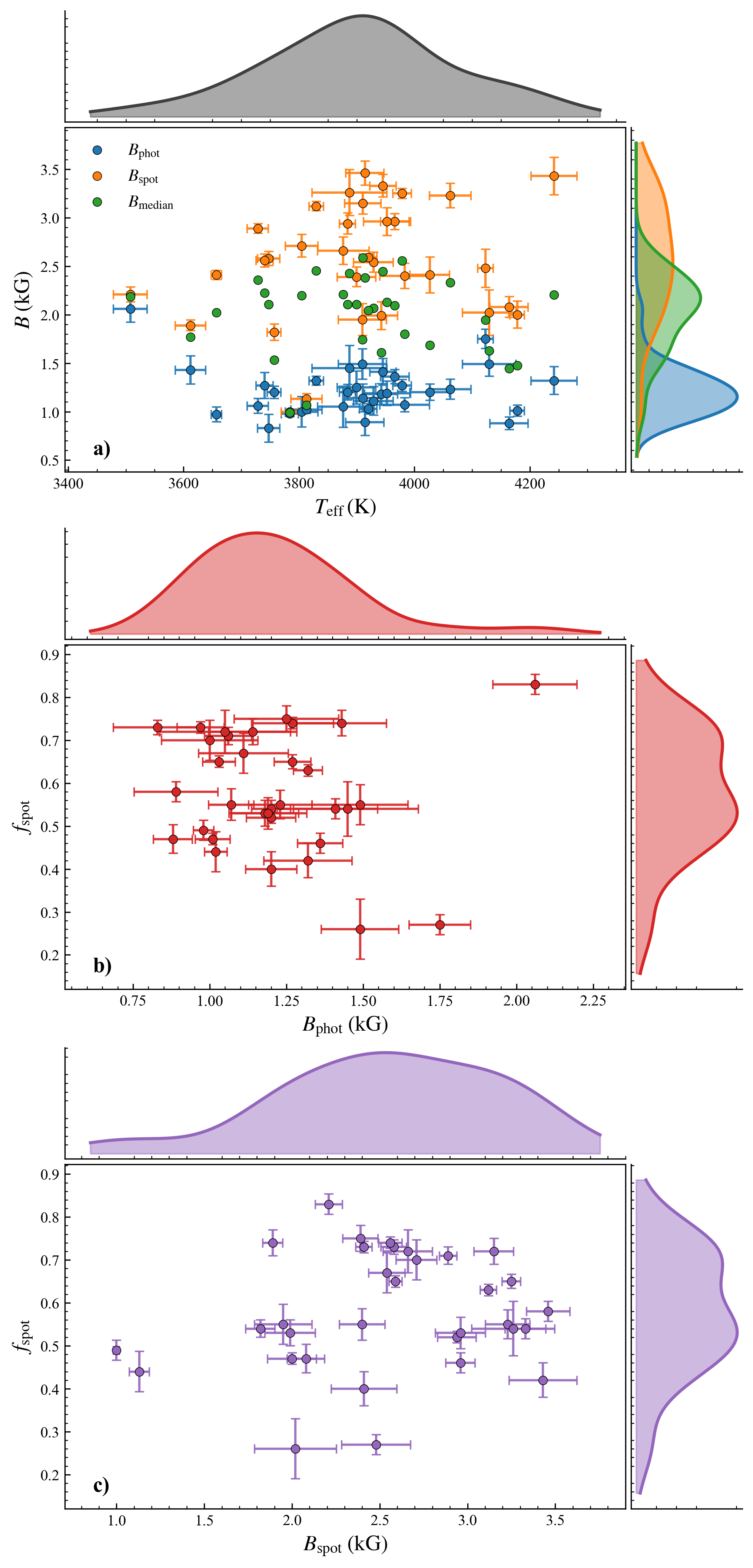}
\caption{{Magnetic field strengths and starspot filling factors for the sample. }{(a)} Effective temperature versus photospheric (\rm $\rm B_{\mathrm{phot}}$), spot ($\rm B_{\mathrm{spot}}$), and surface-averaged ($\rm B_{\mathrm{average}}$) magnetic field strengths, shown as colored points with $1\sigma$ uncertainties. {(b)} $\rm B_{\mathrm{phot}}$ versus $\rm f_{\mathrm{spot}}$. {(c)} $\rm B_{\mathrm{spot}}$ versus $f_{\mathrm{spot}}$. All panels include marginal kernel density estimates (KDEs) along the top and right axes to illustrate the distributions of each quantity; in panel~(a), the right marginals are color-coded to match the three field strength components. The absence of clear monotonic trends in panels~(b) and~(c) is consistent with the low correlation coefficients reported in Section~\ref{sec:results}.
}\label{fig:B_vs_T}
\end{figure}

\subsection{On Surface Gravity Estimates From H- and K-Band Spectra}
Comparing our surface gravities to those reported for the same stars in \citet{lopez2021, lopez-valdivia_igrins_2023, Flores2022}, we find systematically higher values by 0.1–0.3 dex. When we repeat our fits using only the spectral regions adopted in those studies but excluding the $H$-band window, we recover excellent agreement ($\Delta\log g \leq 0.1$ dex, $\Delta T_{\mathrm{eff}} \leq 30$~K). This indicates that the inclusion of the $H$-band drives our solutions toward higher inferred surface gravities. A similar behavior was noted in \citet{perez_paolino_starspots_2025} when combining optical-NIR spectral regions in their fits to high-resolution X-Shooter spectra of PMS, where it was attributed in part to magnetic evacuation allowing starspot emission to emerge from deeper layers. While this effect may contribute here, even with explicit spot modeling we must also consider the wavelength dependence of opacity in cool stars.

Detailed line-formation calculations show that in K–M dwarfs, continuum and line formation in the $H$-band can occur up to $\sim$15~km deeper in the atmosphere than in the $K$-band \citep{perdomo_garcia_opacity_2023}. At these depths, the opacity is higher by a factor of a few, leading to a corresponding increase in gas pressure ($P \propto g\tau\kappa^{-1}$). Because spectroscopic $\log g$ determinations are sensitive to line broadening and thus to atmospheric pressure, the deeper layers probed in the $H$-band naturally yield higher inferred gravities. In this sense, the H-band values are not “wrong” physically, they represent the pressure at the layers being sampled, but they do not directly measure the global $g = GM_\star/R_\star^2$ of the star. Optical spectra, and in the case of IGRINS the $K$-band, provide more faithful probes of the true surface gravity.

With all of these results at hand, we conclude that stars should be placed on the Kiel-Diagram using $\log~g$ values derived from $K$-band spectra only, such as those reported in \citet{lopez2021}, \citet{lopez-valdivia_igrins_2023}, and \citet{Flores2022}, in conjunction with our spot-corrected $HK$-band effective temperatures.

\subsection{Super-Equipartition Magnetic Fields}
A long-standing tension in the literature concerns surface magnetic field strengths that appear to meet or exceed the equipartition value
\begin{equation}
    B_{\rm eq} \simeq \sqrt{8\pi P_{\rm gas}}
\end{equation}
the field at which magnetic pressure balances the local photospheric gas pressure, $P_{gas}$ (e.g., \citealt{Safier1999}). Zeeman broadening studies of T~Tauri stars frequently report mean surface field strengths of $2$--$4$~kG, assuming a single temperature and magnetic field (e.g., \citealt{johnskrull_magnetic_2007}). Similar kG-level fields are found among very active cool stars and M~dwarfs (e.g., \citealt{Shulyak2019}). These measurements have sometimes been interpreted as super-equipartition when compared to photospheric pressures. One solution to this problem that several papers have discussed is the possibility that the magnetic filling factor approaches unity, effectively resulting in entirely magnetized stellar surfaces. 

These field strengths are not direct measurements of a global dipole, but rather averages over all visible magnetic components, convolved with stellar rotation, line formation depths, and instrumental resolution. Complementary techniques such as ZI \citep[e.g.,][]{Basri1992, Shulyak2017}, spectropolarimetry and ZDI \citealt{Donati1997, Donati2009}, and radius inflation constraints from eclipsing binaries \citep[e.g.,][]{Torres2010, Feiden2016} all highlight different facets of the underlying magnetic topology, though each suffers from inherent degeneracies between intrinsic field strength and surface filling factor.

Our two-component analysis provides a realistic and physically motivated resolution by allowing distinct magnetic fields in the warm photosphere and in the cool starspots instead of a single global field. In this picture, the photosphere hosts magnetic field strengths under or near equipartition, allowing gas pressure to dominate or be equalized by magnetic pressure, while the spots, formed where convection is magnetically inhibited, naturally reach higher strengths. Radiative-MHD simulations of sunspots and cool starspots suggest that umbrae can sustain fields exceeding the local equipartition estimate because the gas pressure itself is reduced by strong cooling and partial evacuation, and because the relevant balance is between total (thermal + turbulent + radiative) pressure and magnetic pressure in a multi-component atmosphere (e.g., \citealt{Rempel2012}; see also \citealt{browning_simulations_2008}, \citealt{yadav_explaining_2015} for strong fields in global dynamo models). Thus, kG-level $B_{\rm spot}$ values above a photospheric $B_{\rm eq}$ are not only plausible but expected.

\subsection{Comparison to ZDI Magnetic Field Strengths}
ZDI studies of pre-main sequence stars report a wide range of large-scale surface magnetic field strengths, often dominated by a strong poloidal component, for several stars in our sample. For example, \citet{donati_magnetic_2015} find V819~Tau to host a primarily poloidal field (80–90\% of total energy) with an average unsigned flux of 0.37~kG and peak radial fields of about 0.5~kG within spots. In contrast, our fits indicate a 1.36~kG photospheric field and stronger 3.12~kG spot fields. Similarly, \citet{nicholson_surface_2018} report a mean field of 0.113 kG (0.296~kG peak) for TWA 9A (CD-36 7429A), whereas we recover 1.27~kG in the photosphere and 3.25~kG in the spots. Such large discrepancies likely arise because our method is sensitive to the total magnetic flux, including small-scale regions, while ZDI only recovers the large-scale component. 

While these ZDI results are internally consistent within the MaTYSSE collaboration, several methodological choices likely contribute to the systematic underestimation of total field strengths. ZDI inversions typically adopt optical atomic lines that minimize spot contamination when deriving stellar parameters. This practice biases effective temperatures, surface gravities, and magnetic field strengths toward the unspotted photosphere, even though much of the magnetic flux originates in cooler spots. In reality, the observed profiles are a convolution of contributions from regions with different temperatures and field strengths. By fitting only a single-temperature photosphere, ZDI systematically neglects the magnetic flux from cooler components.

Although MaTYSSE papers acknowledge wavelength-dependent temperature differences and the role of spots, they have yet to test the impact of incorporating two-temperature models or lowering effective temperatures to match spot-corrected or molecular-band-based estimates \citep[e.g.,][]{herczeg2014}. Our two-component fits may provide a complementary way of way of estimating the contrast between magnetic topologies, even if limited to distinguishing only between two average field strengths.

\section{Broader Implications of These Results}
Magnetic activity is now recognized as a major contributor to the observed discrepancies between evolutionary models and empirical constraints from eclipsing binaries, interferometry, and cluster sequences \citep[e.g.,][]{Somers2015,Somers2017,Feiden2016}. Standard stellar structure models, which neglect magnetic fields and surface inhomogeneities, systematically overpredict effective temperatures and underpredict radii for low-mass PMS stars \citep{Torres2010}. In recent years, magnetically inflated models have attempted to reconcile these offsets by incorporating a uniform suppression of convection or an effective spot coverage factor into the stellar interior calculations \citep[e.g.,][]{Jackson2014,Feiden2016,MacDonald2017}. Our results provide direct, empirical constraints on the magnetic topology underlying these effects. This two-component magnetic structure is qualitatively consistent with magneto-convective models in which a modest global field reduces convective efficiency across the surface, while locally strong fields in starspots fully inhibit convection and lower the local gas pressure, allowing for super-equipartition field strengths \citep[e.g.,][]{yadav_explaining_2015}.

The magnitude of $f_{\mathrm{spot}}$ we measure implies a substantial reduction in the bolometric flux, which, when integrated over the stellar surface, naturally explains the cooler $T_{\mathrm{eff}}$ and inflated radii observed in young stars \citep[e.g.,][]{Somers2015,Jackson2014}. Importantly, the separation of $B_{\mathrm{phot}}$ and $B_{\mathrm{spot}}$ offers a physically motivated way to couple surface magnetic maps to interior structure models, rather than assuming a single field strength or uniform surface suppression. Furthermore, these measurements serve as calibration points for evolutionary tracks that include magnetic effects. For example, models such as those of \citet{Feiden2016} could incorporate our empirically determined $f_{\mathrm{spot}}$ and $B$-field component values to reproduce the observed HR diagram positions of young stars without requiring ad hoc parameters. As a result, the application of our two-component field formalism to magneto-convective stellar evolution codes may yield more accurate ages and masses for PMS stars, circumstellar disk and accretion evolution \citep[e.g.,][]{Hillenbrand2008}, and the early dynamical histories of planetary systems.

Beyond stellar structure, these results have direct consequences for exoplanet science. Large spot coverage fractions, especially with $B_{\mathrm{spot}}$ above photospheric equipartition, can significantly alter the apparent stellar spectrum, biasing both radial velocity measurements and transit spectroscopy. For transmission spectroscopy in particular, wavelength-dependent spot and faculae contrasts imprint spurious features in exoplanet atmosphere retrievals if not properly modeled \citep[e.g.,][]{Rackham2018,Zhang2018Trappist}. 

\section{Conclusion}
We have presented the first systematic decomposition of surface magnetic fields in pre–main-sequence stars into separate photospheric ($B_{\mathrm{phot}}$) and starspot ($B_{\mathrm{spot}}$) components using high-resolution $H$- and $K$-band spectra with Zeeman-intensification models. By fitting a two-temperature, two-field model to a carefully vetted subsample of 33 young stars, we find median values of $B_{\mathrm{phot}}\approx1.2$ kG and $B_{\mathrm{spot}}\approx2.56$ kG, with typical spot filling factors of $f_{\mathrm{spot}}\sim0.27$–$0.83$. The photospheric fields seem to cluster below the expected equipartition strength, while the spot fields reach super-equipartition levels in a manner consistent with magneto-convective models and radiative–MHD simulations of spot umbrae. Future work on larger samples will test this hypothesis.

These results provide direct empirical evidence for a structured magnetic topology on the surfaces of young stars, in which cool, magnetically inhibited starspots contain magnetic field strengths factors of a few larger than the photosphere. This picture offers a physically grounded resolution to the long-standing “super-equipartition” problem in Zeeman broadening measurements and links naturally to magnetically inflated stellar evolution models. The high $f_{\mathrm{spot}}$ values we measure imply substantial reductions in $T_{\mathrm{eff}}$ and therefore reduce the level of so-called radius inflation, providing a direct calibration for models that incorporate magnetic effects into pre–main-sequence tracks.

Beyond stellar evolution, our framework has broad implications for exoplanet science. By separating $B_{\mathrm{spot}}$ from $B_{\mathrm{phot}}$, we enable more realistic modeling of spot-induced line distortions in both radial velocity planet searches and transmission spectroscopy of exoplanet atmospheres, mitigating false signals and biased atmospheric inferences.

\begin{deluxetable*}{lcccccccccccc}
\tablecaption{Sample and Fitting Results. \label{tab:obs}}
\tablehead{
\colhead{Star} & \colhead{$f_{\mathrm{spot}}$} & \colhead{$T_{\mathrm{phot}}$ (K)} & \colhead{$T_{\mathrm{spot}}$ (K)} & \colhead{$\log g^{(\text a)}$} & \colhead{$v \sin i$ (km/s)} & \colhead{$v_r$ (km/s)} & \colhead{$r_H$} & \colhead{$r_K$} & \colhead{$B_{\mathrm{phot}}$ (kG)} & \colhead{$B_{\mathrm{spot}}$ (kG)} & \colhead{$\Delta$BIC$^{\text (b)}$} & \colhead{$T_{\mathrm{eff}}$ (K)}}
\startdata
AA~Tau & $0.73 \pm 0.05$ & $4247 \pm 90$ & $3499 \pm 30$ & $4.01 \pm 0.03$ & $11.9 \pm 0.6$ & $27.84 \pm 0.22$ & $0.56 \pm 0.07$ & $0.85 \pm 0.04$ & $0.8 \pm 0.4$ & $2.58 \pm 0.22$ & $205.93$ & $3748 \pm 57$ \\
BP~Tau & $0.70 \pm 0.14$ & $4052 \pm 156$ & $3682 \pm 52$ & $4.31 \pm 0.07$ & $8.5 \pm 0.4$ & $26.37 \pm 0.20$ & $0.92 \pm 0.11$ & $1.20 \pm 0.07$ & $1.0 \pm 0.5$ & $2.7 \pm 0.3$ & $68.61$ & $3805 \pm 85$ \\
CD-36~7429A & $0.65 \pm 0.05$ & $4376 \pm 67$ & $3703 \pm 17$ & $4.50 \pm 0.02$ & $10.68 \pm 0.21$ & $24.53 \pm 0.10$ & $0.14 \pm 0.03$ & $0.05 \pm 0.02$ & $1.27 \pm 0.18$ & $3.25 \pm 0.16$ & $833.28$ & $3979 \pm 48$ \\
DI~Tau & $0.49 \pm 0.07$ & $4011 \pm 36$ & $3494 \pm 34$ & $3.91 \pm 0.05$ & $11.55 \pm 0.24$ & $-4.73 \pm 0.11$ & $0.29 \pm 0.06$ & $0.17 \pm 0.03$ & $0.98 \pm 0.10$ & $1.00 \pm 0.05$ & $195.95$ & $3784 \pm 44$ \\
DN~Tau & $0.54 \pm 0.06$ & $4002 \pm 28$ & $3502 \pm 29$ & $4.08 \pm 0.05$ & $9.18 \pm 0.25$ & $30.74 \pm 0.11$ & $0.37 \pm 0.04$ & $0.39 \pm 0.02$ & $1.20 \pm 0.19$ & $1.82 \pm 0.25$ & $108.41$ & $3757 \pm 36$ \\
DoAr~25 & $0.52 \pm 0.04$ & $4251 \pm 32$ & $3416 \pm 49$ & $4.00 \pm 0.03$ & $14.6 \pm 0.4$ & $-33.98 \pm 0.18$ & $0.49 \pm 0.05$ & $0.68 \pm 0.03$ & $1.20 \pm 0.24$ & $2.9 \pm 0.3$ & $436.60$ & $3884 \pm 42$ \\
IW~Tau & $0.71 \pm 0.06$ & $4115 \pm 90$ & $3529 \pm 32$ & $4.15 \pm 0.04$ & $8.51 \pm 0.18$ & $11.57 \pm 0.08$ & $0.15 \pm 0.03$ & $0.07 \pm 0.02$ & $1.06 \pm 0.23$ & $2.89 \pm 0.15$ & $688.89$ & $3729 \pm 55$ \\
JH~108 & $0.73 \pm 0.04$ & $4003 \pm 35$ & $3500 \pm 17$ & $4.01 \pm 0.02$ & $12.2 \pm 0.3$ & $30.17 \pm 0.13$ & $0.23 \pm 0.03$ & $0.07 \pm 0.01$ & $0.97 \pm 0.23$ & $2.41 \pm 0.14$ & $209.78$ & $3657 \pm 27$ \\
LkCa~14 & $0.47 \pm 0.04$ & $4496 \pm 30$ & $3701 \pm 25$ & $4.50 \pm 0.02$ & $20.8 \pm 0.4$ & $10.75 \pm 0.22$ & $0.16 \pm 0.03$ & $0.10 \pm 0.02$ & $1.01 \pm 0.17$ & $2.0 \pm 0.4$ & $128.20$ & $4178 \pm 37$ \\
LkCa~4 & $0.83 \pm 0.07$ & $4001 \pm 110$ & $3376 \pm 87$ & $3.72 \pm 0.07$ & $27.6 \pm 0.5$ & $8.59 \pm 0.27$ & $0.45 \pm 0.05$ & $0.23 \pm 0.03$ & $2.1 \pm 0.4$ & $2.21 \pm 0.23$ & $69.55$ & $3508 \pm 87$ \\
LkCa~7 & $0.74 \pm 0.04$ & $4246 \pm 67$ & $3501 \pm 23$ & $4.14 \pm 0.06$ & $15.6 \pm 0.4$ & $-7.02 \pm 0.24$ & $0.28 \pm 0.05$ & $0.15 \pm 0.02$ & $1.3 \pm 0.4$ & $2.56 \pm 0.20$ & $142.58$ & $3740 \pm 44$ \\
ROX~27 & $0.53 \pm 0.09$ & $4249 \pm 95$ & $3594 \pm 72$ & $4.15 \pm 0.08$ & $15.6 \pm 0.6$ & $-26.26 \pm 0.30$ & $0.72 \pm 0.08$ & $1.11 \pm 0.06$ & $1.2 \pm 0.3$ & $2.0 \pm 0.4$ & $65.46$ & $3943 \pm 86$ \\
V1075~Tau & $0.27 \pm 0.07$ & $4250 \pm 23$ & $3705 \pm 63$ & $4.50 \pm 0.02$ & $27.9 \pm 0.5$ & $45.43 \pm 0.29$ & $0.30 \pm 0.04$ & $0.20 \pm 0.03$ & $1.75 \pm 0.30$ & $2.5 \pm 0.6$ & $36.77$ & $4123 \pm 41$ \\
V819~Tau & $0.63 \pm 0.04$ & $4249 \pm 28$ & $3500 \pm 30$ & $4.02 \pm 0.03$ & $8.94 \pm 0.25$ & $-5.56 \pm 0.11$ & $0.17 \pm 0.03$ & $0.08 \pm 0.01$ & $1.32 \pm 0.14$ & $3.12 \pm 0.14$ & $354.23$ & $3830 \pm 37$ \\
CI~Tau & $0.40 \pm 0.12$ & $4249 \pm 86$ & $3602 \pm 143$ & $4.19 \pm 0.08$ & $9.8 \pm 0.5$ & $-13.71 \pm 0.24$ & $0.99 \pm 0.09$ & $1.84 \pm 0.08$ & $1.20 \pm 0.25$ & $2.4 \pm 0.6$ & $44.05$ & $4027 \pm 103$ \\
DK~Tau & $0.72 \pm 0.15$ & $4247 \pm 220$ & $3699 \pm 94$ & $4.33 \pm 0.08$ & $14.4 \pm 0.6$ & $42.9 \pm 0.3$ & $0.82 \pm 0.11$ & $1.79 \pm 0.10$ & $1.1 \pm 0.6$ & $2.7 \pm 0.4$ & $29.20$ & $3877 \pm 134$ \\
FV~Tau & $0.26 \pm 0.21$ & $4253 \pm 104$ & $3701 \pm 197$ & $4.36 \pm 0.09$ & $34.0 \pm 0.9$ & $-7.3 \pm 0.5$ & $1.05 \pm 0.15$ & $1.21 \pm 0.08$ & $1.5 \pm 0.4$ & $2.0 \pm 0.7$ & $0.47$ & $4130 \pm 139$ \\
GG~Tau & $0.44 \pm 0.14$ & $4013 \pm 45$ & $3502 \pm 89$ & $4.01 \pm 0.06$ & $9.41 \pm 0.26$ & $13.02 \pm 0.16$ & $0.63 \pm 0.06$ & $0.94 \pm 0.05$ & $1.02 \pm 0.11$ & $1.13 \pm 0.17$ & $79.44$ & $3813 \pm 81$ \\
V*~VV~Sco & $0.72 \pm 0.09$ & $4337 \pm 142$ & $3699 \pm 71$ & $4.50 \pm 0.05$ & $6.8 \pm 0.6$ & $-35.73 \pm 0.21$ & $0.79 \pm 0.10$ & $1.31 \pm 0.09$ & $1.1 \pm 0.4$ & $3.1 \pm 0.3$ & $150.51$ & $3911 \pm 93$ \\
DoAr~15 & $0.67 \pm 0.14$ & $4282 \pm 99$ & $3713 \pm 74$ & $4.47 \pm 0.04$ & $7.3 \pm 0.7$ & $-28.27 \pm 0.22$ & $0.50 \pm 0.11$ & $1.09 \pm 0.10$ & $1.1 \pm 0.4$ & $2.5 \pm 0.3$ & $80.06$ & $3929 \pm 103$ \\
EM*~SR 8 & $0.55 \pm 0.14$ & $4232 \pm 118$ & $3570 \pm 123$ & $4.16 \pm 0.08$ & $25.9 \pm 0.5$ & $-25.1 \pm 0.3$ & $0.36 \pm 0.07$ & $0.20 \pm 0.04$ & $1.5 \pm 0.5$ & $1.9 \pm 0.5$ & $79.94$ & $3910 \pm 126$ \\
VSS~II-23 & $0.54 \pm 0.07$ & $4251 \pm 58$ & $3614 \pm 93$ & $4.28 \pm 0.05$ & $16.5 \pm 0.5$ & $-24.32 \pm 0.28$ & $0.31 \pm 0.04$ & $0.16 \pm 0.02$ & $1.4 \pm 0.4$ & $3.3 \pm 0.3$ & $192.31$ & $3946 \pm 68$ \\
DoAr~24 & $0.55 \pm 0.10$ & $4415 \pm 116$ & $3684 \pm 105$ & $4.28 \pm 0.07$ & $8.7 \pm 0.4$ & $-20.25 \pm 0.19$ & $0.51 \pm 0.09$ & $0.58 \pm 0.06$ & $1.2 \pm 0.3$ & $3.2 \pm 0.4$ & $272.09$ & $4062 \pm 108$ \\
Elia~2-24 & $0.54 \pm 0.19$ & $4222 \pm 207$ & $3513 \pm 183$ & $4.09 \pm 0.11$ & $12.2 \pm 0.9$ & $-34.6 \pm 0.3$ & $1.48 \pm 0.22$ & $2.54 \pm 0.19$ & $1.4 \pm 0.7$ & $3.3 \pm 0.7$ & $57.52$ & $3888 \pm 195$ \\
ISO-Oph~43 & $0.46 \pm 0.07$ & $4256 \pm 70$ & $3515 \pm 91$ & $4.19 \pm 0.04$ & $12.73 \pm 0.26$ & $4.23 \pm 0.13$ & $0.26 \pm 0.03$ & $0.20 \pm 0.02$ & $1.36 \pm 0.22$ & $2.96 \pm 0.25$ & $554.46$ & $3966 \pm 74$ \\
ROXs~12 & $0.65 \pm 0.04$ & $4256 \pm 67$ & $3696 \pm 43$ & $4.35 \pm 0.03$ & $7.75 \pm 0.21$ & $-36.03 \pm 0.09$ & $0.16 \pm 0.03$ & $0.07 \pm 0.02$ & $1.03 \pm 0.16$ & $2.59 \pm 0.11$ & $604.82$ & $3920 \pm 45$ \\
Haro~1-8 & $0.47 \pm 0.10$ & $4467 \pm 94$ & $3716 \pm 89$ & $4.47 \pm 0.06$ & $5.2 \pm 0.5$ & $-22.37 \pm 0.14$ & $0.35 \pm 0.07$ & $0.48 \pm 0.05$ & $0.88 \pm 0.19$ & $2.1 \pm 0.3$ & $139.33$ & $4164 \pm 99$ \\
WLY~2-49 & $0.53 \pm 0.11$ & $4274 \pm 145$ & $3582 \pm 108$ & $4.16 \pm 0.08$ & $7.8 \pm 0.7$ & $-4.21 \pm 0.19$ & $0.81 \pm 0.10$ & $1.17 \pm 0.09$ & $1.2 \pm 0.4$ & $3.0 \pm 0.4$ & $210.88$ & $3953 \pm 122$ \\
DoAr~33 & $0.55 \pm 0.11$ & $4351 \pm 151$ & $3581 \pm 109$ & $4.27 \pm 0.08$ & $10.9 \pm 0.4$ & $-36.45 \pm 0.17$ & $0.50 \pm 0.10$ & $0.80 \pm 0.08$ & $1.07 \pm 0.22$ & $2.4 \pm 0.4$ & $307.72$ & $3983 \pm 130$ \\
V*~V2129~Oph & $0.58 \pm 0.07$ & $4342 \pm 123$ & $3484 \pm 85$ & $4.06 \pm 0.07$ & $13.7 \pm 0.6$ & $-33.01 \pm 0.27$ & $0.55 \pm 0.07$ & $0.64 \pm 0.05$ & $0.9 \pm 0.4$ & $3.5 \pm 0.4$ & $473.86$ & $3914 \pm 99$ \\
ISO-Oph~195 & $0.75 \pm 0.09$ & $4361 \pm 199$ & $3702 \pm 44$ & $4.52 \pm 0.06$ & $17.2 \pm 0.5$ & $2.3 \pm 0.3$ & $0.24 \pm 0.07$ & $0.52 \pm 0.06$ & $1.2 \pm 0.5$ & $2.39 \pm 0.30$ & $95.02$ & $3900 \pm 100$ \\
WSB~63 & $0.74 \pm 0.09$ & $4004 \pm 104$ & $3438 \pm 66$ & $3.91 \pm 0.04$ & $14.3 \pm 0.4$ & $-0.78 \pm 0.13$ & $0.20 \pm 0.04$ & $0.21 \pm 0.02$ & $1.4 \pm 0.4$ & $1.89 \pm 0.17$ & $288.43$ & $3612 \pm 79$ \\
Haro~1-16 & $0.42 \pm 0.12$ & $4496 \pm 114$ & $3791 \pm 138$ & $4.67 \pm 0.08$ & $12.4 \pm 0.7$ & $-23.09 \pm 0.30$ & $0.89 \pm 0.11$ & $1.78 \pm 0.10$ & $1.3 \pm 0.4$ & $3.4 \pm 0.6$ & $62.31$ & $4242 \pm 119$ \\
\enddata
\vspace{0.1cm}
(a)~These surface gravities are unreliable due to being biased towards higher values. See text for a full discussion. (b)~ Difference in BIC between the best-fit unspotted single magnetic field model and the two magnetic-field model.
\end{deluxetable*}
\begin{acknowledgments}
This work used The Immersion Grating Infrared Spectrometer (IGRINS), which was developed under a collaboration between the University of Texas at Austin and the Korea Astronomy and Space Science Institute (KASI) with the financial support of the US National Science Foundation under grants AST-1229522, AST-1702267 and AST-1908892, McDonald Observatory of the University of Texas at Austin, the Korean GMT Project of KASI, the Mt. Cuba Astronomical Foundation and Gemini Observatory. The RRISA is maintained by the IGRINS Team with support from McDonald Observatory of the University of Texas at Austin and the US National Science Foundation under grant AST-1908892. {We are grateful to the many PIs and observers whose efforts in designing, proposing, and carrying out IGRINS programs, as well as the staff that reduced the data and made the archival RRISA resource possible. Their investment of telescope time and expertise created the dataset that enabled this study. We also thank the anonymous referee for their careful and constructive review, which greatly improved the clarity and strength of this work.}
\end{acknowledgments}

\restartappendixnumbering
\appendix
\section{Sample Posterior Parameter Distribution}

\begin{figure*}[hb!]
\centering
\includegraphics[width=1\linewidth]{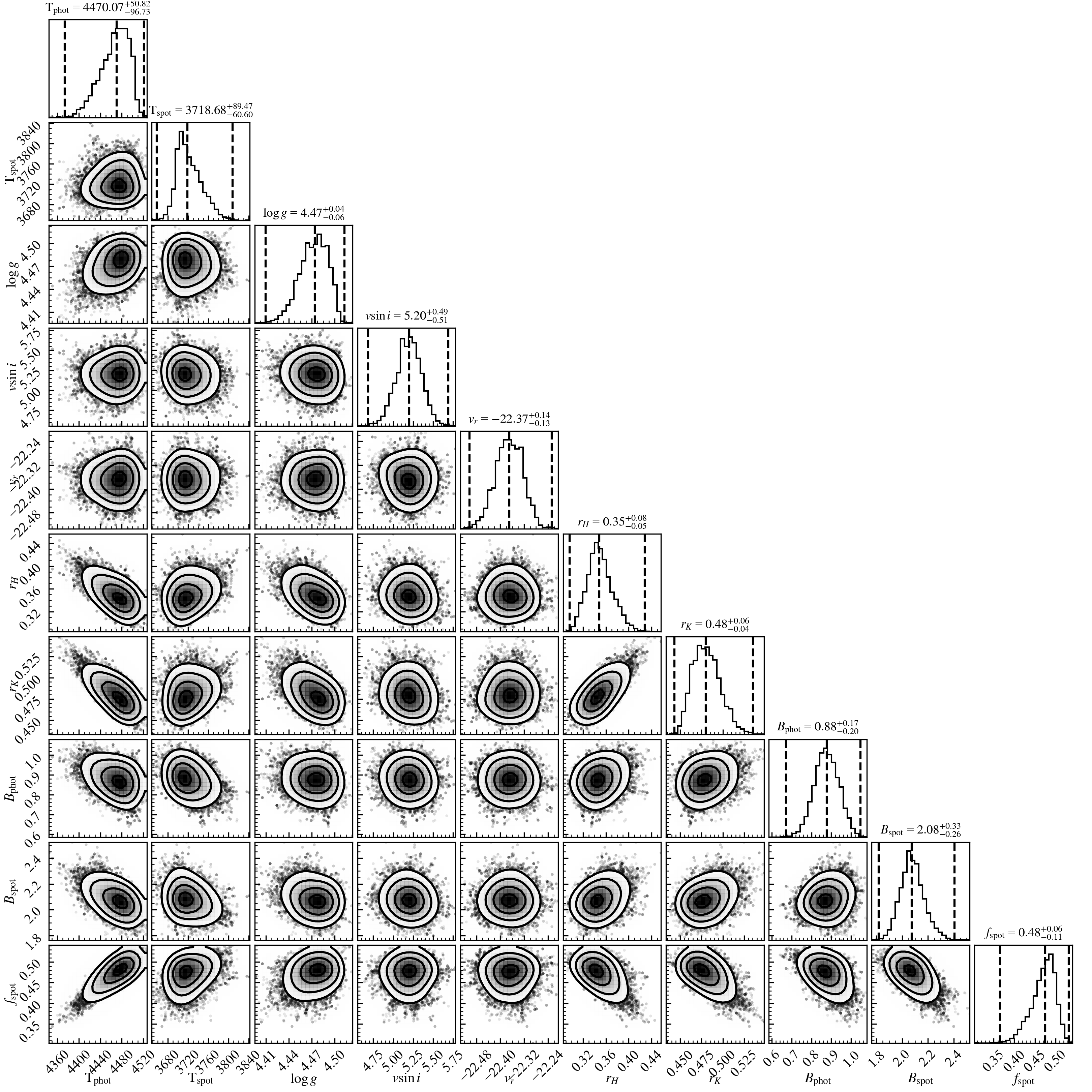}
\caption{{Posterior probability distributions for the fitted parameters of Haro 1-8 derived using the two-component spotted magnetic model.} Vertical lines mark the median and $\pm 3\sigma$ intervals. Off-diagonal panels show the corresponding two-dimensional marginalized distributions for each parameter pair, with contours enclosing the $1\sigma$, $2\sigma$, and $3\sigma$ confidence regions.}\label{fig:posterior}
\end{figure*}

\pagebreak
\bibliography{spots_working.bib}{}
\bibliographystyle{aasjournalv7}

\end{document}